
\input harvmac
\def\rhob{{\rho\kern-0.465em \rho}}

\def\ontopss#1#2#3#4{\raise#4ex \hbox{#1}\mkern-#3mu {#2}}

\setbox\strutbox=\hbox{\vrule height12pt depth5pt width0pt}

\def\strut{\relax\ifmmode\copy\strutbox\else\unhcopy\strutbox\fi}


\nref\rkm{R. Kedem and B.M. McCoy, J. Stat. Phys. 71 (1993),883.}

\nref\rkkmma{R. Kedem, T.R. Klassen, B.M. McCoy and E. Melzer,
Phys. Letts. B304(1993) 263.}

\nref\rkkmmb{R. Kedem, T.R. Klassen, B.M. McCoy and E. Melzer,
Phys. Letts. B 307 (1993) 68.}

\nref\rdkkmm{S. Dasmahapatra, R. Kedem, T.R. Klassen, B.M. McCoy and
E. Melzer, Int. J. Mod. Phys. B (1993) 3617.}
\nref\rdkmm{S. Dasmahapatra, R. Kedem, B.M. McCoy and E. Melzer,
J. Stat. Phys. 74 (1994) 239.}

  \nref\rkmm{R. Kedem, B.M. McCoy, and E. Melzer, The sums of Rogers,
  Schur and Ramanujan and the Bose-Fermi correspondence in $1+1$
  dimensional quantum field theory, hepth-9304056.}

  \nref\rff{B.L. Feigin and D.B. Fuchs, Funct. Anal. Appl. 17 (1983)
  241.}

  \nref\rrc{A. Rocha-Caridi, in {\it Vertex Operators in Mathematics and
  Physics,}
  ed. J. Lepowsky, S. Mandelstam and I.M. Singer. (Springer, Berlin, 1985).}

  \nref\rkp{V.G. Kac and D.H. Peterson, Adv. Math. 53 (1984) 125.}

  \nref\rjm{M. Jimbo and T. Miwa, Adv. Stud. in Pure MAth, 4 (1984) 97.}
  \nref\rciz{A. Cappelli, C. Itzykson and J.-B. Zuber, Nucl. Phys. B 280
  (1987) 445.}
  \nref\rkwa{V.G. Kac and M. Wakimoto, Adv. Math. 70 (1988) 156.}

  \nref\rfel{G. Felder, Nucl. Phys. B317 (1989) 215.}
  \nref\rroga{L.R. Rogers, Proc. London Math. Soc. (Series) 25 (1894) 319.}

  \nref\rsc{I. Schur, Berliner Sitzungberichte 23 (1917) 301.}

  \nref\rrogb{L.J. Rogers, Proc. London Math. Soc. (Series 2) 16 (1917) 315.}
  \nref\rrrh{L.J. Rogers, S. Ramanujan and G.H. Hardy,
   Proc. Cambridge Phil. Soc. 19 (1919) 211.}

  \nref\rlp{J. Lepowsky and M. Primc,{\it Structure of the standard
  modules for the affine Lie algebra $A^{(1)}_1$,} Contemporary
  Mathematics, Vol. 46 (AMS, Providence, 1985). }

  \nref\rter{M. Terhoeven, Lift of the dilogarithm to partition
  identities, BONN-He-92-36 (1992), hepth 9211120.}
  \nref\rkns{A. Kuniba, T. Nakanishi and J. Suzuki, Mod. Phys. Lett. A8
  (1993) 1835.}

  \nref\rkr{J. Kellendonk and A. Recknagel, Phys. Lett. 298B (1993)
  329.}

  \nref\rdas{S. Dasmahapatra, String hypothesis and characters of coset
  CFT's, hepth9305024.}

  \nref\rfeist{B. Feigin and A. Stoyanovsky, Quasi-particles models for
  the representations of Lie algebras and geometry of flag manifold,
  preprint RIMS 942 (1993)}
  \nref\rkrv{J. Kellendonk, M. R{\"o}sgen and R. Varnhagen,
  Int. J. Mod. Phys. A9 (1994) 1009.}

  \nref\rter{M. Terhoeven, Mod. Phys. Lett. A9 (1994) 133.}
  \nref\rmela{E. Melzer, Int. J. Mod. Phys. A9 (1994) 1115.}

  \nref\rmel{E. Melzer, Lett. Math. Phys. 31 (1994) 233.}
  \nref\rdas{S. Dasmahapatra, On State Counting and Characters, hepth 9404116}
  \nref\rber{A. Berkovich, Nucl. Phys. B431 (1994) 315.}

  \nref\rfqa{O. Foda and Y-H Quano, Polynomial Identities of the Rogers
  Ramanujan Type, hepth 9407191.}

  \nref\rfqb{O. Foda and Y-H. Quano, Virasoro character Identities from
  the Andrews-Bailey Construction, hepth 9408086.}

  \nref\rkir{A. Kirillov, Dilogarithm Identities, hepth 9408113.}
  \nref\rwpa{S.O. Warnaar and P.A. Pearce, Exceptional Structure of the
  dilute $A_3$ model; $E_8$ and $E_7$ Rogers-Ramanujan identities,
  hepth-9408136.}

  \nref\rwpb{S.O. Warnaar and P.A. Pearce, A-D-E Polynomial Identities
  and Rogers-Ramanujan Identities, hepth-9411009}

  \nref\rbpz{A.A. Belavin, A.M. Polyakov and A.B. Zamolodchikov,
  J. Stat. Phys. 34 (1984) 763, and Nucl. Phys. B241 (1984) 333.}

  \nref\rgko{P. Goddard, A. Kent and D. Olive, Comm. Math. Phys. 103
  (1986) 105.}

  \nref\rkent{A. Kent, PhD. Thesis, Cambridge U. (1986)}
  \nref\rkwb{V. Kac and M. Wakimoto, Proc. Natl. Acad. Sci. USA 85
  (1988) 4956.}

  \nref\rmatwal{P. Mathieu and M. Walton,
  Prog. Theor. Phys. Supppl. 102 (1990) 229.}

  \nref\rbeoo{M. Bershadsky and H. Ooguri, Comm. Math. Phys. 126 (1989)
  49.}

  \nref\rfeifre{B. Feigin and E. Frenkel, Phys. Letts. B246 (1990) 75.}

  \nref\rdrinsok{V. Drinfeld and V. Sokolov, J. Sov. Math. 30 (1984) 1975.}
  \nref\rmupa{S. Mukhi and S. Panda, Nucl. Phys. B338 (1990) 263.}

  \nref\rfno{B.L. Feigin, T. Nakanishi and H. Ooguri,
  Int. J. Mod. Phys. A7, Suppl. 1A (1992) 217.}

  \nref\rgor{B. Gordon, Amer. J. Math. 83 (1961) 393.}

  \nref\rand{G.E. Andrews, Proc. Nat. Sci. USA 71 (1974) 4082.}
  \nref\rts{M. Takahashi and M. Suzuki, Prog. of Theo. Phys. 48 (1972)
  2187.}
\nref\rbazresa{V.V. Bazhanov and N.Yu. Reshetikhin, Int. J. Mod. Phys.
A4 (1989) 115.}
\nref\rbazresb{V.V. Bazhanov and N.Yu. Reshetikhin, J. Phys. A 23 (1990) 1477.}
\nref\rabf{G.E.Andrews, R.J. Baxter and P.J. Forrester,
J. Stat. Phys. 35 (1984) 193.}
  \nref\rgs{C. Gomez and G. Sierra, Nucl. Phys. B352 (1991) 791.}
  \nref\rkirres{A. Kirillov and N. Reshetikhin in Proc. Paris-Meubon
  Colloquium 1968. World. Scientific. Singapore 1987.}
  \nref\rfyf{H. Frahm, N. Yu, and M. Fowler, Nucl. Phys. B336 (1990)
  396.}
\nref\rnep{L. Mezincescu and R.I. Nepomechie, Phys. Letts. A 246
(1990) 412.}
  \nref\rbgs{A. Berkovich, C. Gomez and G. Sierra, Nucl. Phys. B415
  (1994) 681.}
  \nref\rforbax{P.J. Forrester and R.J. Baxter, J. Stat. Phys. 38 (1985) 435.}
\nref\rbail{W.N. Bailey, Proc. London Math. Soc. (2) 49 (1947) 421;
and 50 (1949) 1.}
\nref\randb{G.E. Andrews, Pac. J. Math, 144 (1984) 267.}
\nref\raab{A.K. Agarwal, G.E. Andrews and D.M. Bressoud, J. Indian
Math. Soc. 51 (1987) 57.}
\nref\rbres{D.M. Bressoud, in {\it Ramanujan Revisited}, G.E. Andrews
et al. eds. Academic Press (1988) 57.}

\nref\rahn{C. Ahn, S-W. Chung and S-H. Tye,  Nucl. Phys. B365 ( 1991) 191.}

  \Title{\vbox{\baselineskip12pt\hbox{BONN-TH-94-28}
  \hbox{ITPSB 94-060}
  \hbox{HEP-TH 9412030}}}
  {\vbox{\centerline{Continued Fractions and}
  \centerline{Fermionic Representations for}
  \centerline{Characters of $M(p,p')$ Minimal models}}}
  \centerline{Alexander Berkovich~\foot{berkovic@pib1.physik.uni-bonn.de}}

  \bigskip\centerline{\it Physikalisches Institut der}
  \centerline{\it Rheinischen Friedrich-Wilhelms Universit{\" a}t Bonn}

  \centerline{\it Nussallee 12}
  \centerline{\it D-53115 Bonn. Germany}
  \bigskip
  \centerline{and}
  \bigskip
  \centerline{ Barry~M.~McCoy~\foot{mccoy@max.physics.sunysb.edu}}

  \bigskip\centerline{\it Institute for Theoretical Physics}
  \centerline{\it State University of New York}
  \centerline{\it  Stony Brook,  NY 11794-3840}
\bigskip
\centerline{\it Dedicated to Prof. Vladimir Rittenberg on his 60th birthday}
\Date{\hfill 12/94}

  \eject

  \centerline{\bf Abstract}

  We present fermionic sum representations of the characters
  $\chi^{(p,p')}_{r,s}$  of the minimal $M(p,p')$ models for all
  relatively prime integers $p'>p$ for some allowed values of $r$ and
$s$. Our starting point is binomial
  (q-binomial) identities derived from a truncation of the state
  counting equations of the XXZ spin ${1\over 2}$ chain of anisotropy
  $-\Delta=-\cos(\pi{p\over p'})$. We use the Takahashi-Suzuki method to
  express the allowed values of $r$ (and $s$) in terms of the continued
  fraction decomposition of $\{{p'\over p}\}$ (and ${p\over p'}$) where $\{x\}$
  stands for the fractional part of $x.$
  These values are, in fact, the dimensions of the  hermitian
  irreducible representations of $SU_{q_{-}}(2)$ (and $SU_{q_{+}}(2)$)
  with
  $q_{-}=\exp (i \pi \{{p'\over p}\})$ (and $q_{+}=\exp ( i \pi {p\over
  p'})).$ We also establish the duality relation $M(p,p')\leftrightarrow
  M(p'-p,p')$ and discuss the action of the Andrews-Bailey transformation
  in the space of minimal models. Many new identities of the
Rogers-Ramanujan type are presented.

  \newsec{Introduction}

  Fermionic representations of conformal field theory characters are q
  series of the form
  \eqn\fermi{\sum_{{\bf m},{\rm restrictions}}q^{{\bf m}{{\bf B}\over
  2}{\bf m}+{\bf}A{\bf m}}\prod_{a=1}^n
  {((1-{\bf B}){\bf m}+{{\bf u}\over 2})_a\atopwithdelims[] m_a}_q}
  where ${\bf m}$ is an $n$ component vector of nonnegative integers
  which may be subject to restrictions in the  sum,
  ${\bf B}$ is an $n\times n$ matrix, ${\bf A}$ and ${\bf u}$ are
 $n$ component vectors,
  and the q binomial
  coefficients (Gaussian polynomials) are defined as
  \eqn\qbin{{n\atopwithdelims[] m}_q={(q)_n\over (q)_m (q)_{n-m}}}
  where
  \eqn\qdef{(q)_n=\prod_{k=1}^n(1-q^k)}
and we note
\eqn\limbin{\lim_{n \rightarrow \infty}{n\atopwithdelims[] m}={1\over
(q)_{m}}.}
  These representations are built from a representation of the space of
  states in terms of $n$ quasi particles which obey a Pauli momentum
  exclusion rule with  momentum ranges that  depend linearly on the number of
  particles $\bf m$ in the state~\rkm-\rkmm. They are in contrast to
  bosonic representations  of the characters~\rff-\rfel~
  based upon a truncation of a bosonic Fock space
  where there is no direct quasi-particle interpretation for the states.
  The bosonic representations are best
  adapted for ultraviolet, short distance properties while the fermionic
  representations are best adapted for infrared, long distance
  properties. The bosonic representations are in general unique whereas
  there are usually several different fermionic representations which
  correspond to the different massive integrable perturbations which
  can be put on the conformal field theory. The equality of the
  fermionic and bosonic representations is a generalization of the
  one hundred year old identities of Rogers-Schur-and Ramanujan ~\rroga-\rrrh.

  The theory of bosonic representations is very well developed. However,
  the corresponding fermionic representations are still under
  investigation. The first such representation discovered was for
  the $Z_N$ parafermionic models~\rlp~ and in the last two
  years a large number of fermionic representations of characters and
  branching functions for affine Lie algebras of integer level have
  been found~\rkm-\rkmm,~\rter - \rwpb.

  However, what is probably the best known example of
  conformal field theory, the $M(p,p')$ minimal models of Belavin,
  Polyakov, and Zamolodchikov ~\rbpz~ is not in general of this type  if
  $p\neq p'+1.$ There are at least two ways to see this:

  1. The coset construction~\rgko~ of fractional
  level~\rkent-\rmatwal.

  Here we represent $M(p,p')$ as
  \eqn\fraco{{(A^{(1)}_1)_1\otimes (A^{(1)}_1)_m\over (A^{(1)}_1)_{m+1}}}
  where the level $m$ is
  \eqn\fracdef{m={p\over p'-p}-2~~{\rm or}~~-{p'\over p'-p}-2.}

  2.  The Hamiltonian reduction method~\rbeoo-\rfeifre.

  The origin of this method is in the paper of Drinfeld and
  Sokolov~\rdrinsok. In this context the characters of $M(p,p')$
  are obtained~\rmupa~
  as the ``residue'' of the character of $(A^{(1)}_1)_m$ where the level is
  \eqn\fracdeftwo{m={p\over p'}-2~~{\rm or}~~{p'\over p}-2.}

  All four of these fractional levels will be seen in the fermionic
  representations presented below.

  The
  bosonic form of this character is the well known result of
  Rocha-Caridi~\rrc~
  \eqn\roccar{{\hat \chi}^{(p,p')}_{r,s}=q^{\Delta^{(p,p')}_{r,s}-c/24}
  \chi^{(p,p')}_{r,s}}
  where
  \eqn\roccar{\chi^{(p,p')}_{r,s}=
  {1\over
  (q)_{\infty}}\sum_{k=-\infty}^{\infty}(q^{k(kpp'+rp'-sp)}-
  q^{(kp'+s)(kp+r)}),}
  the conformal dimensions are
  \eqn\dim{\Delta^{(p,p')}_{r,s}={(rp'-sp)^2-(p-p')^2\over 4pp'}~~~~
  (1\leq r \leq p-1,~1 \leq s \leq p'-1),}
  the central charge is
  \eqn\cen{c=1-{6(p-p')^2\over pp'}.}

  For the minimal unitary model $p'=p+1$ the fermionic
  form of $\chi^{(p,p')}_{r,s}$  has
  given by ~\rkkmmb~ and proven  in~\rber. When $p' \neq p+1$
  fermionic forms have been found for the
  following cases:
  \item{1.}$\chi^{(2,2n+1)}_{1,s}$ for $n=2,3,\cdots$ and all $s$~\rfno.
  The equality of the bose and fermi forms are the original identities
  of Rogers-Schur-Ramanujan~\rroga-\rrrh~ when $n=2$ and are the
  Andrews-Gordon identities~\rgor-\rand~  for $n\geq 3.$
   \item{2.}$\chi^{(p,kp+1)}_{1,kj}$ and $\chi^{(p,kp+1)}_{1,k(j+1)+1}$ for
  $p\geq4,1\leq j\leq p-2$ and $k\geq1$ \rkkmmb~\rfqb;
  \item{3.}$\chi^{(p,kp+p-1)}_{1,j(k+1)}$ and
  $\chi^{p,kp+p-1)}_{1,j(k+1)+k}$ for
  $p\geq 4,1\leq j\leq p-2$ and $k\geq 1$ ~\rfqb;
  \item{4.}$\chi^{ (2n+1,k(2n+1)+2)}_{s,kn+1}$
  and  $\chi^{ (2n+1,k(2n+1)+2)}_{s,(k+1)n+1}$
  for $1\leq s\leq k$~\rkkmmb \rfqb;
  \item{5.}$\chi^{(2n+1,k(2n+1)+2n-1)}_{s,kn+k-1}$
  and $\chi^{(2n+1,k(2n+1)+2n-1)}_{s,(k+1)n+k-1}$ for $1\leq s\leq k$ \rfqb.

  In this paper we generalize these results for characters of $M(p,p')$
  minimal models to all $p$ and $p'$. Our method will be a
  generalization of ~\rber. We will first formulate in sec. 2 a
  fermionic  state counting
  problem for the $M(p,p')$ system with a finite number of states and
  relate it to a bosonic counting formula. We will then generalize these
  fermi-bose counting identities to $q$ deformed identities for the
  finite system which in the infinite limit give identities of fermionic
  forms~\fermi~ with the Rocha-Caridi bosonic character~\roccar.
  We consider separately the cases
  $p'>2p$ in sec. 3 and $p'<2p$ in sec. 4 and conclude with
discussions in sec. 5.

  \newsec{Counting and continued fractions for $p'>2p$.}

  The counting formulae for the general minimal models are obtained by
  an extension of the counting techniques of ~\rber~which were developed
  for the unitary models $p'=p+1.$ This counting problem for $M(p,p')$
  is obtained from the thermodynamic treatment of the XXZ spin chain
  \eqn\hxxz{H_{XXZ}=-\sum_k\left(
  \sigma^{x}_{k}\sigma^{x}_{k+1}+\sigma^{y}_{k}\sigma^{y}_{k+1}
  +\Delta \sigma^{z}_{k} \sigma^{z}_{k+1}\right)}
  where $\sigma^{i}_k~(i=x,y,z)$ are Pauli spin matrices and
  \eqn\dpp{\Delta=-\cos\pi{p\over p'}}
  which was solved by Takahashi and Suzuki in 1972~\rts.

  To present the counting
  solution when $p'>2p$ we define the $n+1$ positive integers $\nu_j$
  from the continued fraction decomposition of $p'/p$
  \eqn\confrac{{p'\over p}=
  (\nu_0+1)+{1\over {\nu_1+{1\over\nu_2+\cdots+{1\over\nu_n+2}}}}.}
For $p'>2p$ we have $\nu_0 \geq 1$ and we will tacitly assume
$\nu_n\geq 0.$ The case $\nu_n=-1$ follows with only a slight change in
notation.
The interval
  \eqn\int{0\leq k \leq \sum_{j=0}^{n}\nu_j}
  is divided into $n+1$ subintervals $(i=1,\cdots,n+1)$
  \eqn\subint{1+t_{i-1}\leq k \leq t_{i}}
  where
  \eqn\takdef{t_i=\sum_{j=0}^{i-1}\nu_j,~~{\rm for}~~i=1,\cdots,n+1~~{\rm
  and}~~t_{0}=-1.}
  We refer to the index $j$ of $\nu_j$ as the ``zone'' index,
  $\nu_j$ as the number of bands in the $j^{th}$ zone, $t_{i+1}$ is
the upper end of the zone $i$, and we say that
  when ~\subint~holds then k lies in the zone $i$ (note that our $t_i$
  is $m_i-1$ of ~\rts) . We also adopt the convention that when
$\nu_n=0$ we shall say that there is a zone $n$ even though that zone
contains no bands. The
  counting problem for the models $M(p,p')$ differs from the counting
  problem of the XXZ chain  only in that in the XXZ chain the number of
  bands in the last $n^{th}$ zone is $\nu_n+2$ instead of
  $\nu_n.$ This reduction of
  the number of quasi-particles in the last zone is sometimes referred
  to as a truncation of the space of states and has been often
used~\rbazresa-\rbazresb in
connection with the RSOS models~\rabf.

  From the $\nu_j$ we define the sets of integers $y_i$
  recursively as
  \eqn\ydef{y_{-1}=0,~~y_0=1,~~y_1=\nu_0+1,
  ~~y_{i+1}=y_{i-1}+(\nu_i+2\delta_{i,n})y_i.}
  We then define what we refer to as the set of allowed string
  lengths $l_k$ in each zone $i$ as
  \eqn\ldef{
l_k=y_{i-1}+[k-1-t_i] y_i
  ~~{\rm for}~~1+t_{i}\leq k \leq t_{i+1 }}
  and we note as examples
  \eqn\stringex{\eqalign{l_k&=k~{\rm for}~0\leq k \leq \nu_0=t_1\cr
  &=1+(\nu_0+1)(k-\nu_0-1)~~{\rm for}~~\nu_0+1\leq k \leq \nu_0+\nu_1=t_2,\cr
  {}&=(\nu_0+1)+[(\nu_0+1)\nu_1+1](k-\nu_0-\nu_1-1)~~{\rm
  for}~~t_2+1 \leq k \leq t_3.\cr}}
  We also define a second set of integers $z_i$
  \eqn\zdef{z_{-1}=0,~~z_0=1,~~z_1=\nu_1,~~
  z_{i+1}=z_{i-1}+(\nu_{i+1}+2\delta_{i+1,n})z_i,}
  and a second set of string lengths ${\tilde l}_k$ by
  \eqn\ltilde{{\tilde l}_k=z_{i-1}+[k-1-{\tilde t}_i]z_i~~{\rm
  for}~~1+{\tilde t}_i\leq k \leq {\tilde t}_{i+1}}
  where ${\tilde t}_i=t_{i+1}-\nu _0.$
  It is clear that the $z_i$ are obtained  from
  the same set of recursion relations as the $y_i$ except that zone zero
  is removed in the partial fraction decomposition of $p'/p.$ The
  removal of this zone zero is equivalent to considering a new XXZ chain
  with a anisotropy $\Delta'=-\cos\pi\{{p'\over p}\}$ where $\{x\}$ denotes
  the fractional part of $x.$

  It may now be proven by induction that
  \eqn\yoverz{{y_{i+1}\over z_i}=(\nu_0+1)+{1\over \nu_1+\cdots {1\over
  \nu_i.}}}
  It is then clear from ~\confrac~ that
  \eqn\pandpp{p'=y_{n+1}~~{\rm and}~~p=z_n.}
  We may also prove by induction that
  \eqn\yzrel{y_{i+1} z_{i-1}-y_i z_i=(-1)^{i-1}.}
  Then taking $i$ to be $n$ we obtain
  \eqn\min{(-1)^{n-1}=y_{n+1}z_{n-1}-y_n z_n=p'z_{n-1}-py_n.}
  From the definition~\dim~ we see that the
  minimal conformal dimension is obtained for the values $r_{\rm min}$ and
  $s_{\rm min}$ which satisfy
  \eqn\mineqn{|p'r_{\rm min}-ps_{\rm min}|=1.}
  and thus from~\min~ that
  \eqn\rsmin{s_{\rm min}=y_n~~{\rm and}~~r_{\rm min}=z_{n-1}.}

  The counting problem is obtained from equations of the thermodynamic
  Bethe's Ansatz of the XXZ chain (eqn (3.9) of ~\rts) by taking the
  $0^{th}$ Fourier component and using the notation that the
nonnegative integers
  $n_j~(m_j)$ are number of particle (hole) excitations of type $j.$
  Thus, with the truncation of the number of particles in the last
  ($n^{th}$) zone from $\nu_n+2$ to $\nu_n$ we consider the following
  generalization of the equation (1.10) of~ \rber~
  \eqn\counting{\eqalign{
  n_k+m_k&={1\over 2}(m_{k-1}+m_{k+1})~~{\rm for}~~1\leq k \leq
  t_{n+1} -1~{\rm and}~k\neq t_i,~i=1,\cdots ,
  n\cr
  n_{k_i}+m_{k_i}&={1\over 2}(m_{k_i-1}+m_{k_i}-m_{k_i+1})~{\rm
  for}~k_i=t_i,~i=1,\cdots,n\cr
  n_{t_{n+1}}+m_{t_{n+1}}&={1\over2}
(m_{t_{n+1}-1}+m_{t_{n+1}}\delta_{\nu_n,0}) \cr}}
where by definition $m_0=L.$ (To simplify our presentation we assume
throughout the rest of this section that $\nu_n\neq 0.$)

  This construction of $M(p,p')$ out of the XXZ model is closely
  related to the representation of $M(p,p')$ in terms of two
  quantum groups~\rgs~(and references therein)$SU_{q_{\pm}}(2)$ where
  \eqn\qpmdef{q_{+}=e^{\pi i p\over p'}~~{\rm and}~~q_{-}=e^{\pi i
  p'\over p}}
  which is clearly related to the fractional levels of~\fracdeftwo. In
  this quantum group construction the string lengths $l_k$ and ${\tilde
  l}_k$ can be interpreted as the dimensions of the hermitian
  representations of $SU_{q_{+}}(2)$ ($SU_{q_{-}}(2)$)~\rkirres-\rbgs.

To proceed further we convert ~\counting into the following partition
problem for $L$
\eqn\partition{\sum_{i=1}^{t_{n+1}}n_i l_i+{m_{t_n+1}\over
2}l_{(t_{n+1}+1)}={L\over 2}}
where $l_{(t_{n+1}+1)}=p'-2y_n.$ Depending on whether $m_{t_n+1}$ is even
or odd we consider two cases which we will refer to as ``e'' and ``o''
in what follows
\eqn\eodef{\eqalign{{}&e~~{\rm is~the~case}~~m_{t_{n+1}}=2n_{(t_{n+1}+1)}\cr
{}&o~~{\rm is~the~case~}~~m_{t_{n+1}}=2n_{(t_{n+1}+1)}+1\cr}}
Given the set $(n_1,n_2,\cdots , n_{(t_{n+1}+1)})$ satisfying
{}~\partition~ we can solve ~\counting~ to obtain the companion set
$\{m_1,m_2,\cdots ,m_{t_{n+1}-1}\}$

  The particles and holes in \counting~ are fermionic
  and thus for each $n_j$ and
  $m_j$ the number of distinct states in the band $j$ is given by the binomial
  coefficient
  \eqn\rbinom{{n_j+m_j \choose m_j}={(m_j+n_j)!\over m_{j}! n_{j}!}.}
  Thus the fermionic counting problem for the
  $M(p,p')$ models is the evaluation
  of
  \eqn\fsum{F(L)_{e,o}=\sum_{m_{t_{n+1}}={\rm even,odd}}
  \prod_{j=1}^{t_{n+1}}{n_j+m_j\choose n_j}}
  where the sum is with respect to all solutions of ~\partition.  Our result is
that for L even
  \eqn\ferbose{F(L)_e=B_{s_{\rm min},s_{\rm min}}(L)}
  and for $L+p'$ even
  \eqn\ferboso{F(L)_o=B_{s_{\rm min},p'-s_{\rm min}}(L)}
  where the bose counting function that of Forrester and Baxter~\rforbax
  \eqn\bos{B_{a,b}(L)=
  \sum_{j=-\infty}^\infty\left[{L\choose {L+a-b\over
  2}-jp'}-{L\choose {L-a-b\over 2}-jp'}\right].}
  These counting formulae will
  be sufficient to give the characters $\chi^{(p,p')}_{r_{\rm min},s_{\rm
  min}}$ and $\chi^{(p,p')}_{r_{\rm min},p'-s_{\rm min}}.$

  To obtain further characters we consider solutions
  to ~\counting~ with inhomogeneous
  terms. Thus if we denote ~\counting~ symbolically as
  \eqn\symcount{{\bf n}={\bf M}\cdot {\bf m}+{L\over 2}{\bf e}_1}
  where $({\bf e}_k)_j=\delta_{j,k}$ we will consider the equations with
  an inhomogeneous term ${\bf u}$
  \eqn\countone{{\bf n}=
 {\bf M}\cdot {\bf m}+{L\over 2}{\bf e}_1+{{\bf u}\over 2}}
  and define from~\fsum~the corresponding Fermi sum $F(l,{\bf u})_{e,o}.$
  In this paper we confine our attention to those inhomogeneous terms
  where the fermionic counting sum is equal to precisely one bosonic sum
  $B_{r,s}(L).$ These inhomogeneous terms are given in terms of what we
  will call a string configuration
  \eqn\inhomo{\eqalign{{\bf u}^{(j)}_k&=
  \delta_{k,j}-\sum_{l=i}^n\delta_{k,t_l}~~{\rm for}~~t_{i-1} < j \leq
  t_i ~{\rm and}~i\leq n\cr
{}&=\delta_{k,j}~{\rm for}~i=n+1\cr}}
  We will say that $j$ is the endpoint of the string. An example is
shown in table 1. We find that only
  one and two string configurations give a single bosonic sum. In
  particular

  1. One string configurations
  \eqn\feroneterm{\eqalign{F(L,{\bf u}^{(j)})_e&=B_{a,s_{\rm
  min}}(L)~~{\rm with}~~L+a+s_{\rm min}~{\rm even}\cr
  F(L,{\bf u}^{(j)})_o&=B_{p'-a,s_{\rm min}}(L)~~{\rm with}~~
  L+p'+a+s_{\rm min}~{\rm even}\cr}}
  with
\eqn\ad{a=l_{j+1}~~{\rm if}~~t_i<j<t_{i+1}~~{\rm and}~~
y_{i+1}-2y_n\delta_{i,n}~~{\rm if}~j=t_{i+1}}
where $l_k$ is the string length~\ldef.

  2. Two string configurations.

The inhomogeneous term is now given as
\eqn\twou{{\bf u}^{(j_1,j_2)}={\bf u}^{(j_1)}+{\bf u}^{(j_2)}}
and we find
\eqn\fertwoterm{F(L,{\bf u}^{(j_1,j_2)})_o=B_{p'-a,b}(L)~~{\rm
with}~L+p'+a+b~{\rm even}}
where a (and b) are defined by~\ad~with $j$ replaced by $j_1$ (and $j_2$).

  We prove these results by a generalization of the fermionic counting
  methods of ~\rber. We will only sketch the prove here. Details will be
  presented elsewhere.

  Define a generating function for the fermionic sum~\fsum~ with
 the $n_j$ and $m_j$ related by~\countone~with
\eqn\inhomoneeded{{\bf u}=\sum_{j=1}^{t_{n+1}}c_j{\bf e}_j}
as
\eqn\gendef{G(x)_i=\sum_{L}x^{L/2}F(L,{\bf u})_i}
where $i=e,o$ and $c_j$ are some integers.

Then following~\rber~ we can evaluate this generating function as
\eqn\genres{G(\theta)_i=G^{(0)}(\theta)_i\prod_{j=1}^{t_{n+1}}\left(
{U_j(\theta)\over U_1^{l_j}(\theta)}\right)^{c_j}}
where $2\cos \theta=x^{-{1\over 2}}$ and
\eqn\gzerodef{\eqalign{{\sin\theta \sin (p'\theta)\over
\sin(y_n\theta)}G^{(0)}(\theta)_i=& \sin(p'-y_n)\theta~~{\rm
if}~~i=e~\cr
=&\sin y_n\theta~~{\rm if}~~i=o\cr}}
and the generalized Chebyshev polynomials $U_j(\theta)$ are
\eqn\cheb{\eqalign{U_j(\theta)=&{\sin(l_{j+1}\theta)\over\sin(y_i\theta)}~~{\rm
for}~~t_i < j < t_{i+1}\cr
=&{\sin ((y_{i+1}-2y_n\delta_{i,n})\theta)\over \sin (y_i\theta)}~~{\rm
for}~~j=t_{i+1}\cr}}
The resulting contour integral expression for $F(L,{\bf u)})_i$ can be
evaluated by means of residue calculus. When ${\bf u}$ is such that
there is one pole which contributes to the integral the above results
are obtained.
  \newsec{Characters for $p'>2p.$}

  The principle of~\rber~ is that the fermi-bose identities of characters
  are obtained by taking the $L\rightarrow\infty$ limit of polynomial
  identities which are obtained from the counting identities by the
  following ``q-deformations''
  \item{1.}
  \eqn\blq{\eqalign{B_{a,b}(L)&\rightarrow
  B_{a,b;r,s}(L,q)=\cr
  {}&\sum_{j=-\infty}^{\infty}\left[q^{j(jp
  p'+rp'-sp)}{L\atopwithdelims[] {L+a-b\over 2}-jp'}_q
  -q^{(jp+r)(jp'+s)}
  {L\atopwithdelims[] {L-a-b\over 2}-jp'}_q \right]\cr }}
   In the limit
  $L\rightarrow \infty$ we have $B_{a,b;r,s}(L,q)\rightarrow
  \chi^{(p,p')}_{r,s}(q).$
  The function $B_{a,b;,r,s}(L,q)$ is that of~\rforbax.

The deformation of the fermionic sums $F(L,{\bf u})$ depends on
which of $n_j$ and $m_j$ (related by ~\countone) are chosen as
independent variables. For the present case we use as independent
variables
\eqn\newvwec{{\tilde{\bf m}}^t
=(n_1,\cdots,n_{\nu_0},m_{\nu_0+1},\cdots, m_{t_{n+1}}).}
Then using~\countone~we define the following $q$ deformation of
$F(L,{\bf u})$
\item{2.}
\eqn\flq{\eqalign{F&(L,{\bf u},q)_{e,o}=\cr
{}&q^{\Delta}\sum_{e,o,{\rm re
strictions}}
q^{{1\over 2}{\tilde{\bf m}}{\bf B}{\tilde {\bf m}}+{\bf A}
{\tilde{\bf m}}}
\prod_{j=1}^{t_{n+1}}{((1-{\bf B}){\tilde {\bf
m}}+L\sum_{l=1}^{\nu_0}{\bf e}_l+{L\over 2}{\bf e}_{\nu_0+1}+f({\bf u}))_j
\atopwithdelims[] {\tilde m}_j}_q}}
where the $e,o$ restrictions are defined below,
 $\Delta$ is chosen so that $F(L,{\bf u},0)=1$ and the nonzero
elements of the matrix  ${\bf B}$ are
  \eqn\bmat{\eqalign{{}&B_{i,j}=2\pmatrix{1&1&1&\ldots&1\cr
			     1&2&2&\ldots&2\cr
			     1&2&3&\ldots&3\cr
			     \vdots&\vdots&\vdots&\ddots&\vdots\cr
			     1&2&3&\ldots&\nu_o\cr}~~{\rm for}~~1\leq
  i,j,\leq \nu_0\cr
  {}&B_{\nu_0+1,j}=B_{j,\nu_0+1}=j~~{\rm for}~~1\leq j \leq
  \nu_0\cr
  {}&B_{j,j}={{\nu_0}\over 2}\delta_{j,\nu_0+1}
  +(1-{1\over 2}\delta_{j,t_i})~~{\rm
  for}~~\nu_0+1\leq j \leq t_{n+1}-1~{\rm and}~2\leq i\leq n\cr
  {}&B_{t_{n+1},t_{n+1}}=1-{1\over 2}\delta_{\nu_n,0}\cr
  {}&B_{j,j+1}=B_{j,j-1}=-{1\over 2}~~{\rm
  for}~j\neq t_i~{\rm and}~ 2\leq i \leq n\cr
 {}&B_{j,j+1}=-B_{j,j-1}={1\over 2}~{\rm for}~j=t_i\cr}}
and
\eqn\fudefn{f({\bf u})={1\over 2}{\bf B u}_++{1\over 2}{\bf u}_{-}}
where we decompose ${\bf u}={\bf u}_{+}+{\bf u}_{-}~{\rm with}$
\eqn\udecop{\eqalign{
({\bf u}_{+})_i&=({\bf u})_i~{\rm for}~i\leq \nu_0~{\rm zero ~otherwise}\cr
({\bf u}_{-})_i&=({\bf u})_i~{\rm for}~i>\nu_0~{\rm zero~otherwise}.\cr}}
The linear terms and the even (odd) restrictions on ${\tilde m}_k$ are
determined by the vector ${\bf u}.$
The restrictions on $m_j$ are uniquely determined from the vector ${\bf
u}$ and the parity of $m_{t_{n+1}}$ by the requirement that the
numerator in the q-binomial coefficients in ~\flq~be an integer.
To describe the restrictions it is convenient to introduce the vector
$v^{(j)}$ with $t_{l_0}<j\leq t_{l_0+1}+\delta_{l_0,n}$ defined as follows
\eqn\vvec{\eqalign{v^{(j)}_k&=(j-k)~{\rm for}~j>k~{\rm and}~k\geq t_{l_0}\cr
v^{(j)}_k&=v^{(j)}_{k+1}+v^{(j)}_{t_{l+1}+1}~~
t_l\leq k<t_{l+1};~~l<l_0 .\cr}}
In terms of $v^{(j)}$ the restrictions can be written as
\eqn\evenres{\eqalign{{\rm~ even~ restriction}:&
m_i=\sum_{j=1}^{t_{n+1}}v_i^{(j)}a_j~{\rm (mod~2)}\cr
{\rm odd~ restriction}:& m_i=\sum_{j=1}^{t_{n+1}}v_i^{(j)}a_j+
v_i^{(t_{n+1}+1)}~({\rm mod~2})\cr}}
with $a_j=1$ if ${\bf u}_j\neq 0$ and $0$ otherwise.

Let us now proceed to define the linear terms. The simplest case is
${\bf u}=0$. In this case ${\bf A}=0$ and we have
\eqn\nolin{\eqalign{F(L,0,q)_e&=B_{s_{\rm min},s_{\rm min};r_{\rm
min},s_{\rm min}}(L,q)\cr
F(L,0,q)_o&=B_{p'-s_{\rm min},s_{\rm min};p-r_{\rm min},s_{\rm
min}}(L,q)\cr}}
Taking $L\rightarrow\infty$ in ~\nolin~we obtain
\eqn\charone{\chi^{(p,p')}_{r_{\rm min},s_{\rm min}}=q^{\Delta}\sum_{{\tilde
m}_i={\rm even}~(i>\nu_0)}
q^{{1\over 2}{\tilde{\bf m}}{\bf B}{\tilde{\bf
m}}}\prod_{j=1}^{\nu_0+1}{1\over (q)_{{\tilde
m}_j}}\prod_{j=\nu_0+2}^{t_{n+1}}{((1-{\bf B}){\tilde{\bf
m}})_j\atopwithdelims[] {{\tilde m}_j}}_q}
and
\eqn\charone{\chi^{(p,p')}_{r_{\rm min},p'-s_{\rm
min}}=q^{\Delta}\sum_{{\tilde m}_i=v_i^{(t_{n+1}+1)}
({\rm mod 2})({\rm for}~i>\nu_0)}
q^{{1\over 2}
{\tilde{\bf m}}{\bf B}{\tilde{\bf
m}}}\prod_{j=1}^{\nu_0+1}{1\over (q)_{{\tilde
m}_j}}\prod_{j=\nu_0+2}^{t_{n+1}}{((1-{\bf B}){\tilde{\bf
m}})_j\atopwithdelims[] {{\tilde m}_j}}_q}

Consider next one string configurations ${\bf u}^{(j)}$ which ends at the
position~j in the zone~i. We say that ${\bf u}^{(j)}$ is an r string if
\eqn\rstrdef{\eqalign{i&>0~~{\rm and}\cr
A_k&=-{1\over 2}{\bf u}_k^{(j)}~~{\rm~for~k~in~an~even~zone}\cr
{}&=0~~{\rm~otherwise}\cr}}
we say that ${\bf u}^{(j)}$ is an s string if
\eqn\sstring{\eqalign{A_k&=-{1\over 2}{\bf u}_k^{(j)}+{\nu_0-j\over
2}\delta_{k,\nu_0+1}\theta(\nu_0-j)~{\rm~for~k~in~an~odd~zone}\cr
{}&=(k-j)~{\rm for ~k~in~zone~zero~and}~0\leq j <\nu_0\cr
{}&=0 ~~{\rm otherwise}\cr}}
where $\theta(x)=1$ if $x>0$ and zero otherwise.
With these definitions we state the following polynomial
identities;
\eqn\onestrpoly{\eqalign{F(L,{\bf u}^{(j)},q)_e&=B_{a,b;r;s}(L,q)\cr
F(L,{\bf u}^{(j)},q)_o&=B_{p'-a,b;p-r,s}(L,q)\cr}}
where for an r string
\eqn\sar{\eqalign{s&=b=s_{\rm min}\cr
a&=l_{j+1}~{\rm if} j\neq t_{i+1}~{\rm and}~y_{i+1}-2y_n\delta_{i,n}~{\rm
if}~j=t_{i+1}\cr
r&={\tilde l}_{j+1-\nu_0}~{\rm if}j\neq t_{i+1}~{\rm and}~z_i-
2z_{n-1}\delta_{i,n}~{\rm if}~j=t_{i+1}\cr}}
and for an s string
\eqn\sas{\eqalign{s&=b=l_{j+1}~{\rm if}~j\neq t_{i+1}~{\rm and}~
y_{i+1}-2y_n\delta_{i,n}~{\rm if}~j=t_{i+1}\cr
r&=r_{\rm min},~~~a=s_{\rm min}.\cr}}
We illustrate these results for an $s$ string in table 1 and for an
$r$ string in table 2 .

Finally we consider the two string configurations with
\eqn\twoconf{{\bf u}={\bf u}^{(j_1)}+{\bf u}^{(j_2)}~~{\rm and}~~{\bf A}={\bf
A}^{(1)}+{\bf A}^{(2)}}
where ${\bf u}^{(j_1)}$ is an $r$ string and ${\bf u}^{(j_2)}$ is an
$s$ string. Then we have the following polynomial identity
\eqn\twopolyinent{F(L,{\bf u},q)_o=B_{a,b;r,s}(L,q)}
where
\eqn\moretwo{\eqalign{ s&=b=l_{i_2+1}~{\rm if}~j_2\neq
t_{i_2+1}~~{\rm and}~~y_{i_2+1}-2y_n\delta_{i_2,n}~{\rm if}~j_2=t_{i_2+1}\cr
r&=p-{\tilde l}_{j_1+1-\nu_0}~{\rm if}~j_1\neq t_{i_1+1}~~{\rm and }
{}~~p-z_{i_1}+2z_{n-1}\delta_{i_1,n}~{\rm if}~j_1=t_{{i_1}+1}\cr
a&=p'-l_{j_1+1}~{\rm if}~ j_1\neq t_{i_1+1}~~{\rm and}~~
p'-y_{i_1+1}+y_n\delta_{i_1,n}~{\rm if}~j_1=t_{i_1+1}.\cr }}
The corresponding character identities are obtained by taking the
limit $L\rightarrow\infty$ with the help of ~\limbin.
Note in particular that for two string
configurations we only obtain polynomial identities for the odd
case. The possibility of character identities for two strings in the
even case is considered in sec. 5.

The proofs of these results is given by generalizing the methods of
{}~\rber. The details will be published elsewhere.
\newsec{Characters for $p'<2p.$}

  When $p'<2p$ we have $\nu_0=0$ and the counting problem~\counting~
  ceases to make sense because the inhomogeneous term is located in the
  zone zero and now there are no variables in zone zero. In the XXZ
chain~\hxxz~ we see from ~\dpp that the regime $p<p'<2p$ corresponds
to $\Delta>0$ and since the XXZ chain has the symmetry
$H_{XXZ}(\Delta)=-H_{XXZ}(-\Delta)$ the counting problem for $\pm \Delta$
  are the same. Thus we consider the relation
  between $M(p,p')$ and $M(p'-p,p')$ which is obtained by the
  transformation
  $q\rightarrow q^{-1}$ in the finite $L$ polynomials $F(L,{\bf u},q)$
  and $B_{a,b;r,s}(l,q).$ This transformation is a manifestation of
the two related  constructions with the fractional
levels~\fracdef~and~\fracdeftwo. However to implement the
transformation it is mandatory that the $L\rightarrow \infty$ limit be taken
only in the final step.

  Consider first $B_{a,s,;r,s}^{(p,p')}(L,q)$~\blq~where we have made
the dependence on $p$ and $p'$ explicit. Then if we note that from the
  definition~\qbin~
  \eqn\qtrans{{n+m\atopwithdelims[]
  m}_{q^{-1}}=q^{-nm}{n+m\atopwithdelims[] m}_q}
  we find
  \eqn\btrans{B_{a,s;r,s}^{(p.p')}(L,q^{-1})=
q^{-{L^2\over 4}} q^{({a-s\over 2})^2}
B_{a,s;a-r,s}^{(p'-p,p')}(L,q).}

Similarly we consider $q\rightarrow q^{-1}$ in $F(L,q)$ and find
\eqn\ftrans{F(L,{\bf u},q^{-1})_i=q^{-{L^2\over 4}-\Delta}\sum_{{\rm
i-restrictions}}  q^{-{1\over
2}{\bf m}{\bf M}{\bf m}+{\bf m}{\bf A}'} \prod_{j=1}^{t_{n+1}}
{((1+{\bf M}){\bf m}+{{\bf u}\over 2}+
{L\over 2}{\bf e}_1)_j\atopwithdelims[] m_j}_q}
where the i-restrictions are given in sec. 3, $\Delta$ is the
normalization constant,
${\bf M}$ is the matrix of coefficients of~ \counting~and ~\symcount~
(where the antisymmetric elements may be set equal to zero,
and
\eqn\aprime{\eqalign{{\bf A}'_k&={\bf u}_k-{\bf
A}_k\theta(k-\nu_o)+{\nu_0-j_1\over
2}\delta_{k,\nu_0+1}\theta(\nu_0-j_1)\cr
{\bf u}&={\bf u}^{(j_1)}+{\bf u}^{(j_2)}~~{\rm where}~~j_1<j_2\cr}}
This mapping of ${\bf A}\rightarrow {\bf A}'$ means that the parity
rules~\rstrdef-\sstring~ change such that the r rule becomes the s rule and
{\it vice
versa}. With this definition an r (s) string remains an r (s) string
however in this case the $r$ string may be in zone zero.
Since the $L$ dependent factor in the transform of both
$B_{a,s;r,s}^{(p,p')}(L,q)$ and $F(L,{\bf u},q)$ are the same we
derive
\eqn\newform{q^{\tilde \Delta}\sum_{{\bf m},\rm restrictions}q^{-{1\over 2}{\bf
m}{\bf M}{\bf m}+{\bf m}{\bf A'}}
\prod_{j=1}^{t_{n+1}}{((1+{\bf M}){\bf m}+{{\bf u}\over 2}+
{L\over 2}{\bf e}_1)_j\atopwithdelims[]
m_j}_q=B_{a,s,a-r,s}^{p'-p,p'}(L,q).}
Making use of~\limbin~ we take the $\L\rightarrow\infty$ limit and
obtain the fermionic representations for the characters of
$M(p'-p,p').$ As an example we note
\eqn\enexam{q^{\tilde \Delta}\sum_{{m_i=even}}q^{-{1\over 2}{\bf m}{\bf
M}{\bf m}}{1\over (q)_{m_1}}\prod_{j=2}^{t_{n+1}}{((1+{\bf M}){\bf
m})_j\atopwithdelims[] m_j}_q=
\chi^{p'-p,p'}_{s_{\rm min}-r_{\rm min},s_{\rm min}}.}
{}From the physical point of view the map $M(p,p')\rightarrow
M(p'-p,p')$ can be viewed as a particle-hole transformation.

We conclude this section by noting that the polynomials
$B_{a, s;a-r,s}^{p'-p,p'}(L,q)$ introduced here have an explicit
connection with the polynomials  $D^{(k)}_L(s,a,a\pm 1)$ of Forrester
and Baxter (equation 2.3.17 of~ \rforbax). Ignoring an overall constant
the relation is
\eqn\forbaxrel{\eqalign{B_{a,s;a-r,s}^{p'-p,p'}(L,q)&=
D_L^{(k)}(s,a,a+1)~{\rm if~ the~the ~r~string~ends~in~an~odd~zone}\cr
{}&=D^{(k)}_L(s,a,a-1)~{\rm if~the~r~string~ends~in~an~even~zone}}}
and
\eqn\forbaxk{\eqalign {k=\left[{ap\over p'}\right]&= r~{\rm
{}~if~the~r~string~ends~in~an~odd~zone}\cr
{}&=r-1~{\rm if~the~r~string~ends~in~an~even~zone}}}
where $[x]$ denotes the integer part of x. Thus we may say that the
results of this paper provide the fermionic q-series form of the path counting
version of the Rogers-Ramanujan identities proven in ~\rforbax.
\newsec{Discussion.}

In the treatment of the characters given above we started with
$M(p,p')$ for $p'>2p$ and used the duality transformation on the
finite $L$ polynomials to compute the polynomials and the characters
for $M(p'-p,p')$. There is, however, another relation between different
models which was found in several special cases by Foda and
Quano~\rfqb. In this approach we start with the finite $L$ polynomials for
the model $M(p,p')$ and add one more zone to the continued fraction
decomposition by means of the Andrews-Bailey
transformation~\rbail-\rbres. This transformation produces the characters
of the related minimal model with ${\tilde p'}>2{ \tilde p}$
but does not produce the
finite $L$ polynomials. We have checked that our results
are consistent with the Andrews-Bailey transformation. More
explicitly, if one starts , for instance, with $p,p',r=l_j,s=s_{\rm
min}$ and takes $k-1$ steps along the so-called Bailey chain then one
obtains a new model with ${\tilde p}=p',~{\tilde p'}=p+kp',~{\tilde
s}={\tilde l}_j+kl_j,~{\tilde r}=s_{\rm min}.$
It is tempting to interpret the Andrews-Bailey
transformation in terms of a renormalization group flow which connects
different minimal models.

We also note that the duality transformation appears to formally
connect the unitary minimal model $M(p,p+1)$ with the model
$M(1,p+1).$ This is indeed a correct transformation on the finite
lattice. However as pointed out in ~\rabf~ the limit $L\rightarrow
\infty$ needed for the characters does not exist unless further
factors of L are removed. This then leads to the parafermionic
characters of Lepowsky and Primc~\rlp~ which are of the form
of~\fermi~with ${\bf u}=\infty$ where there are additional
restrictions on the summation variables  which may be interpreted as
$Z_N$ charges. This is a sort of restriction which is quite different
from the even (odd) restrictions found above.

This remark leads one to ask whether or not there are other choices of
the independent variables which can be made in the fermionic deformation
procedure  of sections 3 that will also lead to valid character
formulae of some other models. One natural candidate is the model of
fractional level parafermions~\rahn~where one would speculate that all
of the $n_j$ should be chosen as independent variables. But if this is in fact
correct it should be just one of many possible truncations that lead
to fermionic sum formula.  We also, point out that the fermionic
representations of characters contain information about integrable
off critical deformations of conformal field theory. Further
discussion of these questions will be pursued
elsewhere.

 Finally we remark that there are more general relations between fermi
sums and the  bosonic Rocha-Caridi character formulae than the ones
studied in this paper. For example, in this paper we have restricted
our attention to character formulae obtained from a fermionic
counting problem which is solved in terms of a single boson counting
formula ~\bos. A particularly interesting generalization of this are
two string configurations for $m_{t_{n+1}}$ even. Then the
fermionic counting function $F(L,{\bf u})_e$ is not solved by the
single boson problem as in ~\fertwoterm. Nevertheless as long as one of the
two inhomogeneous terms is in the last zone a computer study reveals
that when we take the $L\rightarrow \infty$ limit we still have the
single term character identity for $p'>2p$
\eqn\twoeven{\lim_{L\rightarrow\infty}F(L,{\bf u}^{(j_1)}+{\bf
u}^{(j_2)},q)_e=\
\chi^{(p,p')}_{r,s}(q)}
where $j_1$ is an $r$ string and $j_2$ is an $s$ string with
\eqn\twoevenrs{\eqalign{s&=l_{i_{2}+1}~{\rm if} j_2\neq t_{{i_2}+1}~{\rm and}~
y_{{i_2}+1}-2y_n\delta_{i_2,n}~{\rm if}~j_2=t_{{i_2}+1}\cr
r&={\tilde l}_{j_1+1-\nu_0}~{\rm if}~j_1\neq t_{i_1+1}~{\rm
and}~z_{i_1}-2z_{n-1}\delta_{i_1,n}~{\rm if}~j_1=t_{i_1+1}.\cr}}
Let is now recall that the $s$ and $r$ strings are associated with
$SU_{q_+}(2)$ and $SU_{q_-}(2)$ respectively. Then it is natural to
interpret the breakdown of ~\twoeven~ which occurs when neither of the
strings is in the last zone as a manifestation of the ``Drinfeld
twist'' type of interaction between the two quantum groups.
 In general the relation of fermionic sums to
bosonic sums is a matrix relation where one bosonic Rocha- Caridi
character is given in terms of a linear combination of fermionic
sums. Many examples of this linear combination phenomena have been
seen~\rkkmma-\rdkkmm,~\rfqb. The investigation of this general theory
for the models $M(p,p')$ will be the subject of future investigations.

\bigskip

{\bf Table1.}

We illustrate the rules for the restrictions on $m_j$, the linear terms,
and the vector of inhomogeneous terms ${\bf u}$ of a single $s$ string
for the case $p=9,p'=31$ where $\nu_0=2, \nu_1=2, \nu_2=2.$ Here
$r_{\rm min
}=2$ and $s_{\rm min}=7.$
The positions 1 and 2
are in zone 0, 3 and 4 are in zone 1 and 5 and 6 are in zone 2.
We indicate by $j$ what we
mean by the position of the inhomogeneous term. The difference in
this definition for $j$ in the interior of the zone and $j$ at the
boundary of the zone at 2 and 4 are clearly seen.
We have shown the seven positions of the string when $r=r_{\rm min}$ for
the character where $m_6$ is even.


\vbox{\tabskip=0pt

\offinterlineskip
\halign to 100pt{\strut# \quad &\hfil #\quad&\hfil #\quad&\hfil #\quad&\hfil
#\quad & \hfil #\quad&
\hfil #\quad& \hfil#\quad&
\hfil #\quad&\hfil #\quad& \hfil #\quad&\hfil #\quad& \hfil #\quad&
 \hfil#\quad& \hfil #\quad \cr

&$j$&$m_3$&$m_4$&$m_5$&$m_6$&$u_1$&$u_2$&$u_3$&$u_4$&$u_5$&$u_6$&
 linear term &           $ r$& $ s$\cr
&0& o& e& e& e& 0&$-1$& 0&$-1$& 0& 0&$n_1+2n_2+(2m_3+m_4)/2$ & 2&  1\cr
&1& o& e& e& e& 1&$-1$& 0&$-1$& 0& 0&  $n_2+(m_3+m_4)/2$     & 2&  2\cr
&2& o& e& e& e& 0& 0& 0&$-1$& 0& 0&        $m_4/2 $        & 2&  3\cr
&3& o& e& e& e& 0& 0& 1&$-1$& 0& 0&      $(-m_3+m_4)/2$    & 2&  4\cr
&4& e& e& e& e& 0& 0& 0& 0& 0& 0&         none           & 2&  7\cr
&5& o& o& e& e& 0& 0& 0& 0& 1& 0&         none           & 2& 10\cr
&6& o& e& o& e& 0& 0& 0& 0& 0& 1&         none           & 2& 17\cr}}

{\bf Table 2.}

We illustrate the rules for the restrictions on $m_j$, the linear terms,
and the vector of inhomogeneous terms ${\bf u}$ of a single $r$ string
for same case considered in table 1.
We have shown the four positions of the string when $r=r_{\rm min}$ for
the character where $m_6$ is even.


\vbox{\tabskip=0pt

\offinterlineskip
\halign to 125pt{\strut# \quad &\hfil #\quad&\hfil #\quad&\hfil #\quad&\hfil
#\quad & \hfil #\quad&
\hfil #\quad& \hfil#\quad&
\hfil #\quad&\hfil #\quad& \hfil #\quad&\hfil #\quad& \hfil #\quad&
 \hfil#\quad& \hfil #\quad&# \cr

&$j$&$m_3$&$m_4$&$m_5$&$m_6$&$u_1$&$u_2$&$u_3$&$u_4$&$u_5$&$u_6$&
 linear term &           $ r$& $ s$&\cr
&3& o& e& e& e& 0& 0& 1&$-1$& 0& 0&       none           & 1&  7&\cr
&4& e& e& e& e& 0& 0& 0& 0& 0& 0&         none           & 2&  7&\cr
&5& o& o& e& e& 0& 0& 0& 0& 1& 0&         $-m_5/2$       & 3&  7&\cr
&6& o& e& o& e& 0& 0& 0& 0& 0& 1&         $-m_6/2$       & 5&  7&\cr}}

  \bigskip
  \noindent
  {\bf Acknowledgments.}
  We wish to thank O. Foda, W. Nahm, V. Rittenberg, M. R{\" o}sgen and
  M. Terhoeven for interest and discussions and G. Andrews, Y. Quano,
A. Recknagel, and S.
  Warnaar for most useful correspondence.
  This work was partially supported by NSF
  grant DMR 9404747.

\vfill

\eject
\listrefs

\vfill\eject

\bye
\end